\documentclass[12pt,preprint]{aastex}

\makeindex
\def\lsim{\;\lower4pt\hbox{${\buildrel\displaystyle <\over\sim}$}\;}

\usepackage{graphicx}
\usepackage{lscape}
\usepackage{multirow}
\newcommand{\va}{Alfv\'{e}n speed }

\shorttitle{Coronal Shock Strength and SEPs}
\shortauthors{Shen, Wang, Ye, Zhao, Gui and Wang}

\begin{document}
\title{Strength of coronal mass ejection-driven shocks near the Sun and its importance in predicting solar energetic particle events}

\author{Chenglong Shen$^{1,2}$,
Yuming Wang$^{1,3,*}$
, Pinzhong Ye$^{1,2}$, X. P. Zhao$^{4}$, Bin Gui$^{1}$, and S. Wang$^{1}$}
\affil{
$^{1}$CAS Key Laboratory of Basic Plasma Physics, School of Earth \& Space Sciences., University of Science \& Technology of China, Hefei, Anhui 230026, P. R. China. (ymwang@ustc.edu.cn, clshen@mail.ustc.edu.cn)}
\affil{$^{2}$State Key Laboratory of Space Weather, Chinese Academy of Science, Beijing, 100080, P. R. China}
\affil{$^{3}$Department of Computational and Data Sciences, George Mason University, Fairfax, VA 22030, USA}
\affil{$^{4}$W. W. Hansen Experimental Physics Laboratory, Stanford University, Stanford, CA 94305, USA}
\affil{($^{*}$ Author for correspondence)}
\begin{abstract}
Coronal shocks are an important structure but without direct observations 
in solar and space physics. The strength of shocks plays a key role in shock-related 
phenomena, such as radio bursts, SEP generation and so on. This paper will 
present an improved method of calculating \va and shock strength near the Sun. 
In the method, observations as many as possible rather than one-dimensional 
global models are used. Two events, a relatively slow CME on 2001 September 15 
and a very fast CME on 2000 June 15, are selected to illustrate the calculation process. 
The calculation results suggest that the slow CME drove a strong shock 
with Mach number of 3.43$\sim$4.18 while the fast CME drove a relatively 
weak shock with Mach number of 1.90$\sim$3.21. 
This is consistent with the radio observations that a stronger and longer 
decameter-hectometric (DH) type II radio burst is found during the first event 
and a short DH type II radio burst during the second event. 
Particularly, the calculation results explain the observational fact that 
the slow CME produced a major solar energetic particle (SEP) event while 
the fast CME did not. Through the comparison between the two events, the 
importance of shock strength in predicting SEP events is addressed.
\end{abstract}

\section{Introduction}
Shock waves in the corona and interplanetary (IP) medium are usually generated
when the speeds of coronal mass ejections (CMEs) exceed
the local magnetosonic wave speed. Shocks particularly interest researchers 
since they are not only associated with violent eruptions from the Sun, but 
also may have strong capability of accelerating particles and therefore lead 
to many notable consequences which possibly influence the geo-space significantly. 
For example, gradual solar energetic particle (SEP) events,
one of the major space weather effect, are formed as the flux of 
energetic particles near the Earth exceeding a threshold. 
Most such energetic particles are thought to be generated at 
CME/interplanetary CME (ICME)-driven shocks 
\citep{Cane_etal_1981, Reames_1999, Cane_Lario_2006}. 
Type II radio bursts, another significant phenomenon in the corona and IP space, 
are excited by energetic electrons which also generally accelerated by shocks 
\citep[e.g.,][]{Nelson_Melrose_1985}. 
Accelerating particles is a distinct ability of shocks that could be rated by the 
shock strength \citep[e.g.,][]{Jones_Ellison_1991}. 
Except those IP shocks arriving at 1 AU and recorded by in-situ spacecraft, 
many shocks of interest, e.g., the coronal shocks within about 5 solar radii, 
are far away from the observers and only have remote measured data. 
Thus how to accurately deduce shock strength is an inescapable step to understand 
the physical processes involved in shock acceleration of particles and other related phenomena.

Up to now, all coronal shocks are identified based on remote sense data. 
They could be diagnosed, for example, in white-light images of corona from 
Large Angle Spectrometric Coronagraph (LASCO, \citealp{Brueckner_etal_1995}) 
on board Solar and Heliospheric Observatory (SOHO) as done by \citet{Vourlidas_etal_2003}, 
and in UV spectra observations from SOHO Ultraviolet Coronagraph Spectrometer 
(UVCS, \citealp{Kohl_etal_1995}) \citep[e.g.,][]{Raymond_etal_2000, Mancuso_etal_2002, 
Raouafi_etal_2004, Ciaravella_etal_2005, Ciaravella_etal_2006}. So far, tens 
coronal shocks have been identified in UVCS spectra.
However, the best signature/indicator of coronal and IP shocks 
is still the decameter-hectometric (DH) to kilometric type II radio bursts 
(call type IIs for short) which have been known for decades. 
Type IIs provide the information of plasma density at shocks, and seem
to be unavoidable observations to estimation of shock speed and strength \citep[e.g.,][]{Mancuso_etal_2002, Mann_etal_2003, Vrsnak_etal_2004, Warmuth_Mann_2005}.

The strength of a shock could be described by the fast-mode
magnetosonic Mach number $M_{ms}=(V_{sh}-V_{sw})/V_{ms}$, where
$V_{sh}$ is the speed of a shock, $V_{sw}$ is the background solar
wind speed, $V_{ms}=\sqrt{\frac{V_{a}^{2}+C_{s}^{2}+|V_{a}^{2}-C_{s}^{2}|}{2}}$ 
is the fast-mode magnetosonic speed by assuming that the wave vector 
is along the magnetic field lines, $V_a$ is the \va, and $C_s$ is the sound
speed. The \va is defined as $V_{a}=\frac{B}{\sqrt{\mu \rho}}$
where $B$ is the strength of the background magnetic field and
$\rho$ is the background plasma density. 
Thus the values of $V_{sh}$, $V_{sw}$, $V_a$ and $C_s$ are required to calculate shock strength.
$V_{sh}$ can be estimated according to the propagation of CMEs in LASCO field of view (FOV), 
which will be addressed in the next section. 
$V_{sw}$ is generally about 150 km/s at 5 $R_s$, the solar radius, and
even smaller below that height \citep{Sheeley_etal_1997}. Therefore, 
in most cases, $V_{sw}$ could be ignored for a shock within 5 $R_s$  
because the speeds of shock-associated CMEs may reach several 
hundreds or even thousands kilometers per second at 2 $R_s$. $V_a$ involves
magnetic field and plasma density, and $C_s$ is related to temperature. 
As to temperature, an isothermal atmosphere is a good
approximation because the coronal temperature is typically at the order of
one million Kelvin throughout the IP space. 
Finally, two parameters, magnetic field and plasma density, are left unresolved. 

Due to the lack of direct observations of magnetic field and plasma density, 
(quasi-)one dimensional global models of magnetic field and density are widely 
used by researchers. For magnetic field, the simplest 1-D model of magnetic 
field is described as $B(r)=B_s(\frac{R_s}{r})^2$, based on the magnetic field 
conservation $B(r)r^2=const$, in which $B_s=2.2$ G is the magnetic
field strength at solar surface $R=R_s$. 
To integrate some features of active regions, 
the model is sometimes modified by being superimposed a magnetic dipole for active
regions as adopted by \citet[e.g.,][]{Mann_etal_2003}. 
For plasma density, there are also some 1-D global models, 
such as the Saito model \citep{Saito_etal_1977}, 
Newkirk model \citep{Newkirk_1961} and combinations of them 
\citep[e.g.,][]{Gopalswamy_etal_2001a}. 
Combined with type II dynamic spectra, these density models 
could be used to deduce the height and speed of shocks. 
The fatal flaw of these models is also obvious. 
These models are highly ideal, and can resolve the variation 
neither in the plane perpendicular to the radial direction nor in temporal,
which is probably significant as seen in section 3 for the variation of 
magnetic field strength and \citet{Cho_etal_2007} for density.

In this paper, we will present an improved method of 
calculating coronal \va and shock Mach number above 2 $R_s$ 
through investigating two CME events. 
In the method, a more realistic model is used to extrapolate the coronal magnetic field 
based on photospheric measurements, DH type II observations are analyzed to obtain the density 
at shocks, and CME snapshots recorded by coronagraphs are 
adopted to estimate the height information.
The two events selected to illustrate the calculation process are a fast CME 
without a major SEP event and a relatively slow CME causing a major SEP event, respectively. 
In this paper, a major SEP event is defined following NOAA's criterion that 
the peak intensity of proton flux with energy $>$10 MeV observed at the near 
Earth space exceeds 10 pfu ($particles/(cm^{2}-s-ster)$), which is also widely used 
by many other researchers.
As will be seen, the calculation results are well consistent with observations. 
Particularly, the two studied events are right to show the importance of shock strength in predicting SEP events. 
The two events are counter examples for the picture that faster CMEs produce 
larger SEP events. 
Our derived shock Mach numbers for the two events just explain the unusual phenomena. 
The next section will introduce the calculation process step by step. 
In section 3, the two events are presented to further illustrate the method. 
A comparison of the consequences between the two events with respect of SEP intensity is given in section 4. 
Finally, in section 5, we summarize the paper and make a brief discussion.

\section{Our method}

\paragraph{Step 1: Calculation of the local plasma density 
by using the dynamic spectra of DH type IIs.}
Solar radio bursts of type II are indicative of shocks
propagation in the corona and heliosphere accompanied by the
electron acceleration \citep{Malitson_etal_1973,
Wild_McCready_1950, Holman_Pesses_1983}. 
They are thought to be produced in the upstream plasma rather than in the
denser downstream plasma and their frequency reflects the local electron 
density \citep{Dulk_etal_1971, Holman_Pesses_1983}. 
Therefore type IIs can be used to deduce the density of the ambient plasma 
that shocks are passing through. Type IIs usually have fundamental (F) and harmonic (H)
components. The fundamental component is emitted at the place
where the wave frequency is equal to the local electron plasma
frequency. From the frequency of the type IIs fundamental
(F) component, we can calculate the local electron number density
through $N_{e}=(\frac{f_{pe}(Hz)}{8.98\times10^3})^2(cm^{-3})$.
Since the presence of multiple species, the following relationship between
the electron number density and the plasma density in the solar
corona suggested by \citet{Priest_1982} is used to calculate the
plasma density
\begin{equation}
\rho=1.92\widetilde{\mu} m_{p}N_e=1.92\widetilde{\mu}
m_{p}(\frac{f_{pe}(Hz)}{8.98\times10^3})^2(cm^{-3})\ ,
\label{eq_rho}
\end{equation}
where $m_p$ is proton mass with the mean molecular weight
$\widetilde{\mu}=0.6$\citep{Priest_1982}. 

In this paper, we study the ambient conditions and the coronal 
shocks only in the DH frequency range because 
(1) the DH frequency range (from about 1 to 14 MHz) corresponds
to the heliocentric distance about 2--10
$R_s$, where CME-driven shocks are thought to be most efficient for
energetic particles generation \citep{Kahler_1994,
Cliver_etal_2004},
(2) the DH type IIs are thought to be produced at the noses of shocks, not 
like the metric type IIs whose source regions are not clear
\citep[e.g.,][]{Gary_etal_1984,Maia_etal_2000,Gopalswamy_etal_1997,Cho_etal_2007,Robinson_Stewart_1985, Klein_etal_1999,Gopalswamy_Kaiser_2002}. 
The space-borne instrument, Wind/Waves \citep{Bougeret_etal_1995}, just
monitors the radio emissions in this frequency range, and could be
used to deduce the local plasma density we wanted.

\paragraph{Step 2: Calculation of the heliocentric distance where 
DH type II radio bursts emitted and the speed of shocks at that time by using 
SOHO/LASCO images.}
SOHO/LASCO C2 and C3 cameras give the time sequence of the CME
propagation from 1.5 to 30 solar radii.
It can be used to measure the height-time (HT) curve of a CME, 
and read the projected heliocentric distance
and speed of the CME leading front at any given time. 
Following \citet{Gopalswamy_etal_2005} and \citet{Cho_etal_2005}, 
we assume that: (1) DH type II radio
bursts are generated at the noses of CME-driven shocks, (2) the
shock standoff distance is relatively small near the Sun, and can
be therefore ignored. For metric type IIs, many 2-D radio images
show that metric type II sources are behind the CME fronts
\citep{Robinson_Stewart_1985, Klein_etal_1999}. A possible
explanation is that the shocks resulting in metric type IIs are
located at the flank of CMEs due to their expansion
\citep[e.g.,][]{Gopalswamy_Kaiser_2002}. However,
type IIs with longer wavelength are thought to originate ahead of
the CME fronts because CMEs propagate forward and shocks will be
driven ahead of themselves. Thus the first assumption, 
which is also mentioned in the step 1, is reasonable. The 
second assumption is also acceptable since there
are some evidence that the CME bright leading edges are probably
shock fronts \citep[e.g.,][]{Vourlidas_etal_2003,
Ciaravella_etal_2006}. Under these assumptions, CME leading
fronts may act as a proxy of shocks and therefore be related to
DH type IIs. Besides, we assume that CMEs propagate radially, and
correct every measured height, $H_{measure}$, in 2-D LASCO images to 
heliocentric distance, 
$H_{heliocentric\ distance}$, by the equation
\begin{equation}
H_{heliocentric\ distance}=\frac{H_{measure}}
{\sqrt{(\cos{\theta}\sin{\phi})^{2}+\sin^{2}{\theta}}}\
,\label{eq_modi}
\end{equation}
where $\theta$ is the latitude and $\Phi$ is the longitude of the
source region of a CME, which is determined by examining 
SOHO/EIT (Extreme Ultraviolet Imaging Telescope, \citealp{Delaboudiniere_etal_1995}) movies. 
Then we fit the HT scatter plot with a second-order polynomial to get the real CME trajectory.
To assess how good the approximation is, we extrapolate the fitted curve down to the
solar surface, $H_{heliocentric\ distance}=1R_s$, where a CME originates, and compare the
extrapolated CME onset time $T_{OL}$ with the observed onset time
$T_{OE}$ defined as the first appearance of the CME-related
eruptive activity in EIT images. A small difference between
$T_{OE}$ and $T_{OL}$ indicates a good approximation since the
cadence of EIT data is merely about 12 minutes.

\paragraph{Step 3: Calculation of the background magnetic field strength 
at the shock by the CSSS model.}
Up to now, there are no observations of 3-D magnetic field in the corona. 
Most information of coronal magnetic field comes from extrapolation with the aid of models, 
such as the potential field source surface model (PFSS, 
\citealp{Schatten_etal_1969, Altschuler_Newkirk_1969}), 
the potential field-current sheet model (PFCS, \citealp{Schatten_1971}), 
the horizontal current-source surface model (HCSS, 
\citealp{Zhao_Hoeksema_1992, Zhao_Hoeksema_1994a}), 
the horizontal current-current sheet model (HCCS, \citealp{Zhao_Hoeksema_1994b}),
the current sheet-source surface model (CSSS, 
\citealp{Zhao_Hoeksema_1995,Zhao_etal_2002a}), 
and so on. Not like the 1-D global models mentioned in the introduction, 
these models are more robust and complicated, 
in which measured photospheric longitudinal magnetic field is input as the bottom boundary. 
The extrapolated magnetic fields are usually well match the observed large-scale features 
in other wavelengths \citep{Schrijver_etal_2005}.

In this paper, the mature CSSS model is used to extrapolate the coronal magnetic filed. 
The bottom boundary is adopted from the WSO (Wilcox Solar Observatory) 
synoptic charts\footnote{http://wso.stanford.edu/synopticl.html}, which are
assembled from individual magnetograms observed over a Carrington rotation.
Previous works by Zhao and his colleagues have shown that the
extrapolated magnetic field can approximately reproduce the
observed variation of magnetic field in latitude and values at 1
AU \citep{Zhao_Hoeksema_1995, Zhao_etal_2002a}. In our study, the
strength of the magnetic field at shocks is an average of the
extrapolated magnetic field over a circle region with a radius of
$10^\circ$ at the height, where the DH type II emitted, 
above the CME source region. 
One may notice that the extrapolated magnetic field is a kind of average 
over a Carrington rotation and shall not exactly reflect the state at the time of interest. 
However, since what we studied is related to a large scale phenomenon 
and what we wanted is non-CME-disturbed magnetic field, 
we think the averaged state will not significantly distort the reality, 
nevertheless we would like to take an error of 20 percent of magnetic filed strength 
into account to make our calculation results more reliable. 

\paragraph{Step 4: Calculation of the \va and fast-mode magnetosonic Mach number.}
From Step 1 -- 3, the background plasmas density ($\rho$), the height and speed
of the shock ($V_{sh}$), and the magnetic field strength ($B$) are deduced independently.
As have been discussed in the third paragraph of
Introduction, we will ignore the background solar wind speed and
assume an isothermal corona with $T=1.4$ MK that is corresponding
to a constant $C_s$ of 180 km/s. Therefore, the \va and the
fast-mode magnetosonic Mach number indicating the shock strength
can be calculated.

\section{Two Cases}
\subsection{2001 September 15 event}
\begin{figure*}[tb]
\begin{center}
\includegraphics[width=\hsize]{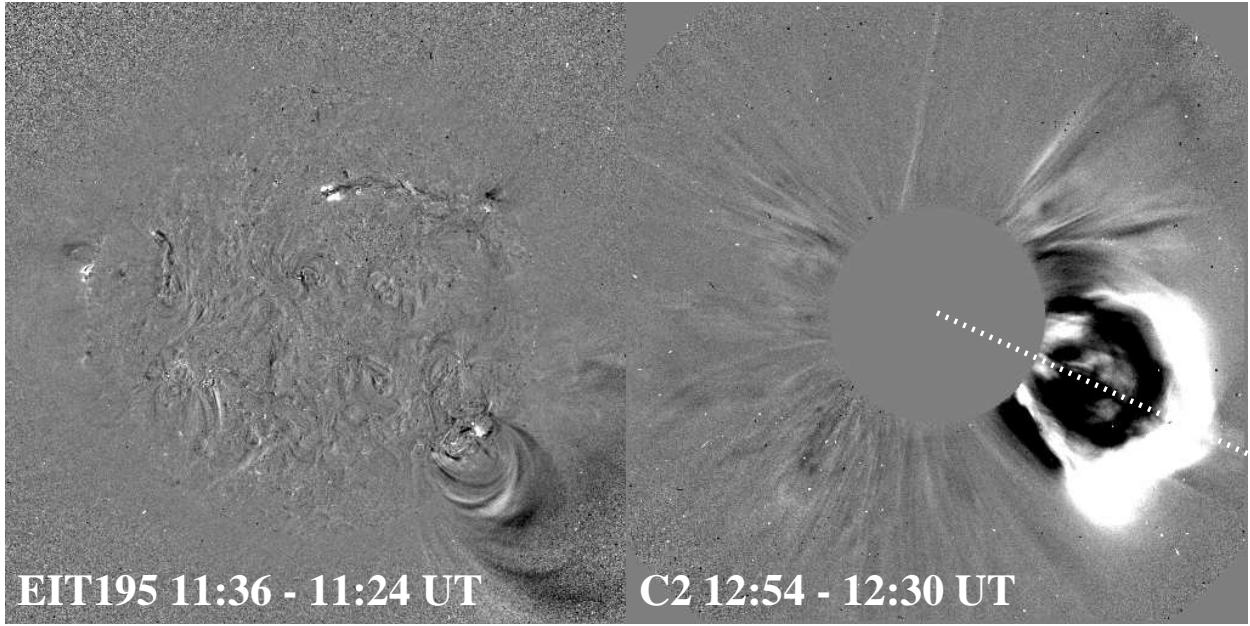}
\caption{The EIT and LASCO running-difference images of 2001 September
15 event. The dashed line indicates the direction along which 
the heights of the CME front were measured.} \label{fg_1_1}
\end{center}

\end{figure*}
\begin{figure*}[tb]
\begin{center}
\includegraphics[width=.9\hsize]{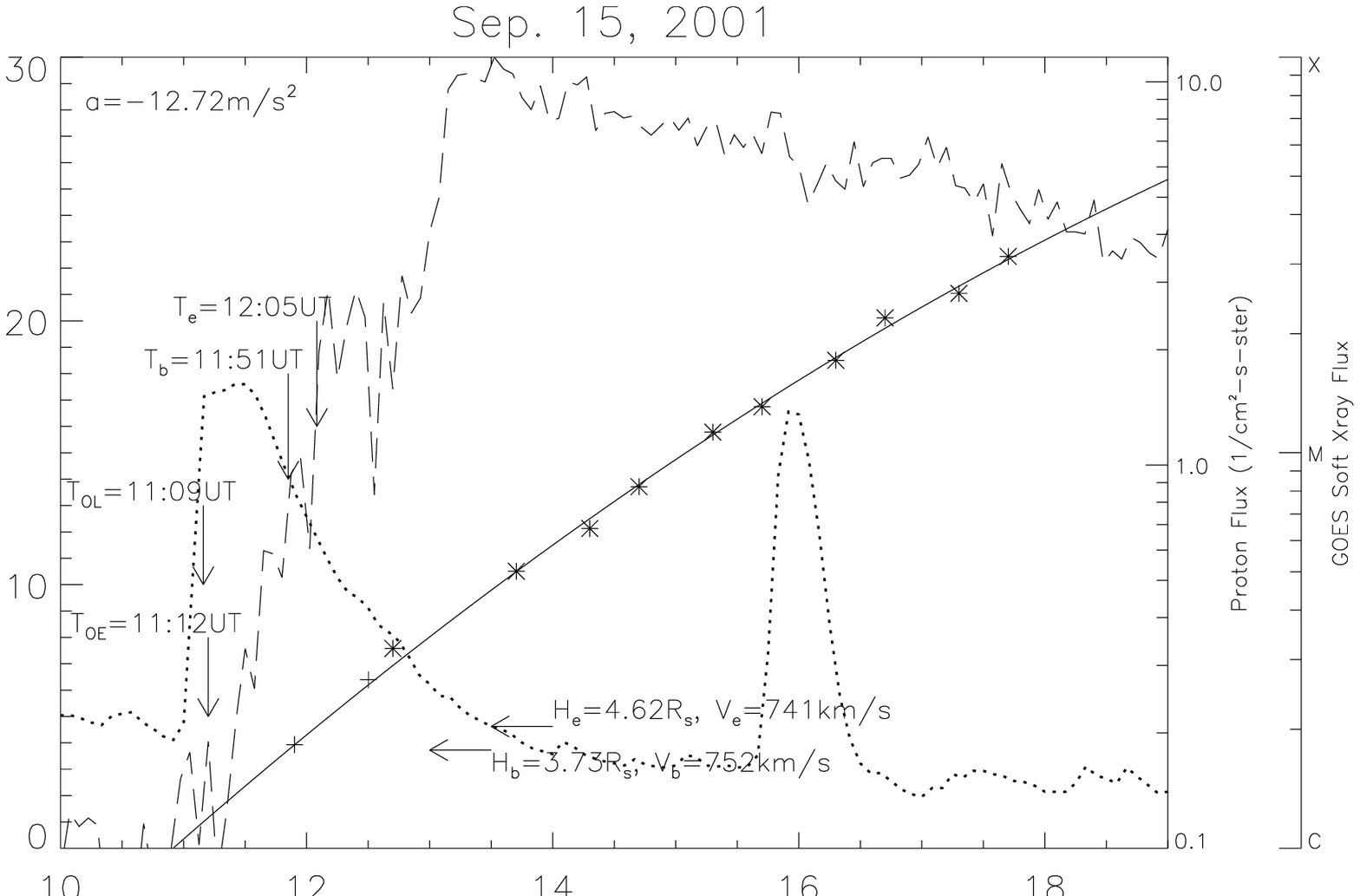}
\caption{The heliocentric height-time plot of 2001 September 15
CME. The cross and asterisk symbols denote the C2 and C3
measurements, respectively. The solid line is the
second-order polynomial fitting line. $T_{OL}$ denotes the extrapolated
CME onset time. $T_{OE}$ is the observed CME onset time in EIT 
images. $T_b$ denotes the begin time of the associated DH type II,
and $H_b$ and $V_b$ denote the height and speed of the CME front
at that time. $T_e$ denotes the end time of the DH type II, and $H_e$
and $V_e$ are the corresponding height and speed of the CME front.
The dashed profile shows the flux of energetic protons with energy larger 
than 10 MeV observed by GOES. A one-hour flight time of protons from 
the Sun to 1 AU has been deducted from the proton flux profile by 
assuming zero scattering in the interplanetary space. The dotted line 
denotes the GOES soft X-ray flux at the wavelength of 1.0 -- 8.0 \AA.} \label{fg_1_2}
\end{center}
\end{figure*}

Figure \ref{fg_1_1} shows the running difference images from
SOHO/EIT and LASCO for this CME. The CME was first observed
by LASCO/C2 at 11:54 UT on September 15, 2001.
It looks like a bright bubble traveling along the southwest direction in the C2
FOV as shown in the right panel of Figure \ref{fg_1_1}. 
The EIT image (left panel) shows that the source
region of this CME is at about S27W48, above which a large EUV
dimming and opening of loops could be seen. 
EIT images and the soft X-ray profile from GOES satellite 
(the dotted line in Figure \ref{fg_1_2}) indicate that the CME was associated
with an M1.5 flare. The significant eruptive signatures in the source region 
first appeared in the 11:12 UT ($T_{OE}$) EIT image.  
By combining the images taken by C2 and C3 cameras, 
we measure the HT profile of the leading
front of this CME until it faded into background. 
The direction of the measurement is chosen artificially, but mainly along the
center of the major part of the CME as indicated by the dashed line in Figure \ref{fg_1_1}. 
The measured heights are further corrected by Equation \ref{eq_modi}, and denoted
by cross (from C2 measurements) and asterisk (from C3 measurements) symbols
in Figure \ref{fg_1_2}. Through a second-order polynomial fitting, the
heliocentric distance of the CME front as a function of time is
obtained. The acceleration is estimated as $-12.72$ m/s$^{2}$,
and the CME speed decreased from 772 km/s at 2 $R_s$ to 674
km/s at 10 $R_s$. From the fitting curve in Figure
\ref{fg_1_2}, the extrapolated onset time, $T_{OL}$, of the CME is
about 11:09 UT. The difference between $T_{OL}$ and $T_{OE}$ is
only 3 minutes indicating that this fitted HT profile is
a good approximation to the reality.

\begin{figure*}[tb]
\begin{center}
\includegraphics[width=.9\hsize]{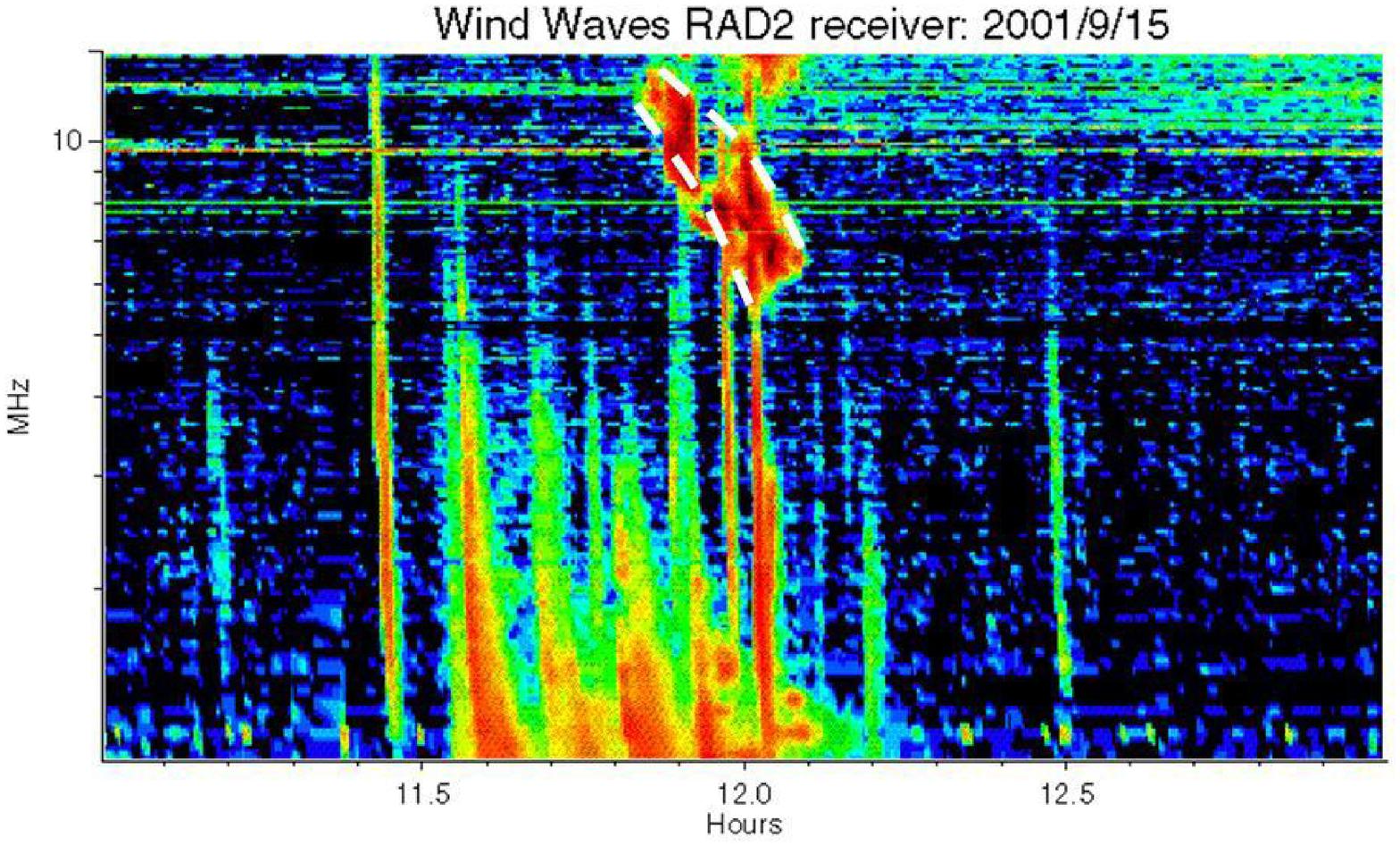}
\caption{The Wind/Waves observations during 11:00 -- 13:00 UT on
September 15, 2001. The dashed lines denote the frequency range of
the fundamental component of a DH type II.}
\label{fg_1_3}
\end{center}
\end{figure*}

This CME was accompanied with a clear DH type II radio burst observed
by the Wind/Waves as exhibited in Figure \ref{fg_1_3}. Since there were no other 
faster CMEs occurring near the CME of interest, the association between the
DH type II and the CME is definite.
The beginning and end times of this DH type II event have been marked by
$T_b$ and $T_e$ in Figure \ref{fg_1_2}. 
The main emission from 11:51 UT ($T_b$) to 12:05 UT ($T_e$) is the fundamental,
and the harmonic just appear at the very top of the detectable frequency
range near the end of the event (M. L. Kaiser 2006, private communication).
Since the fundamental frequency is diffused in a range, two white dashed
lines are used to denote the upper and lower limits of it.
At the beginning of the event, the frequency covers the range from 11.06
to 13.07 MHz approximately, which are corresponding to the plasma
density range from 2.91$\times10^{-18}$ to 4.06$\times10^{-18}$ g/cm$^{3}$.
Both the values are used in the succedent calculations of $V_a$, $V_{ms}$,
and $M_{ms}$ for showing the upper and lower limits of them.
Correspondingly, the
heliocentric distance and speed of the CME are independently estimated as 3.73 $R_s$
and 752 km/s at $T_b$, and 4.62 $R_s$ and 741 km/s at $T_e$.

\begin{figure*}[tb]
\begin{center}
\includegraphics[width=.9\hsize]{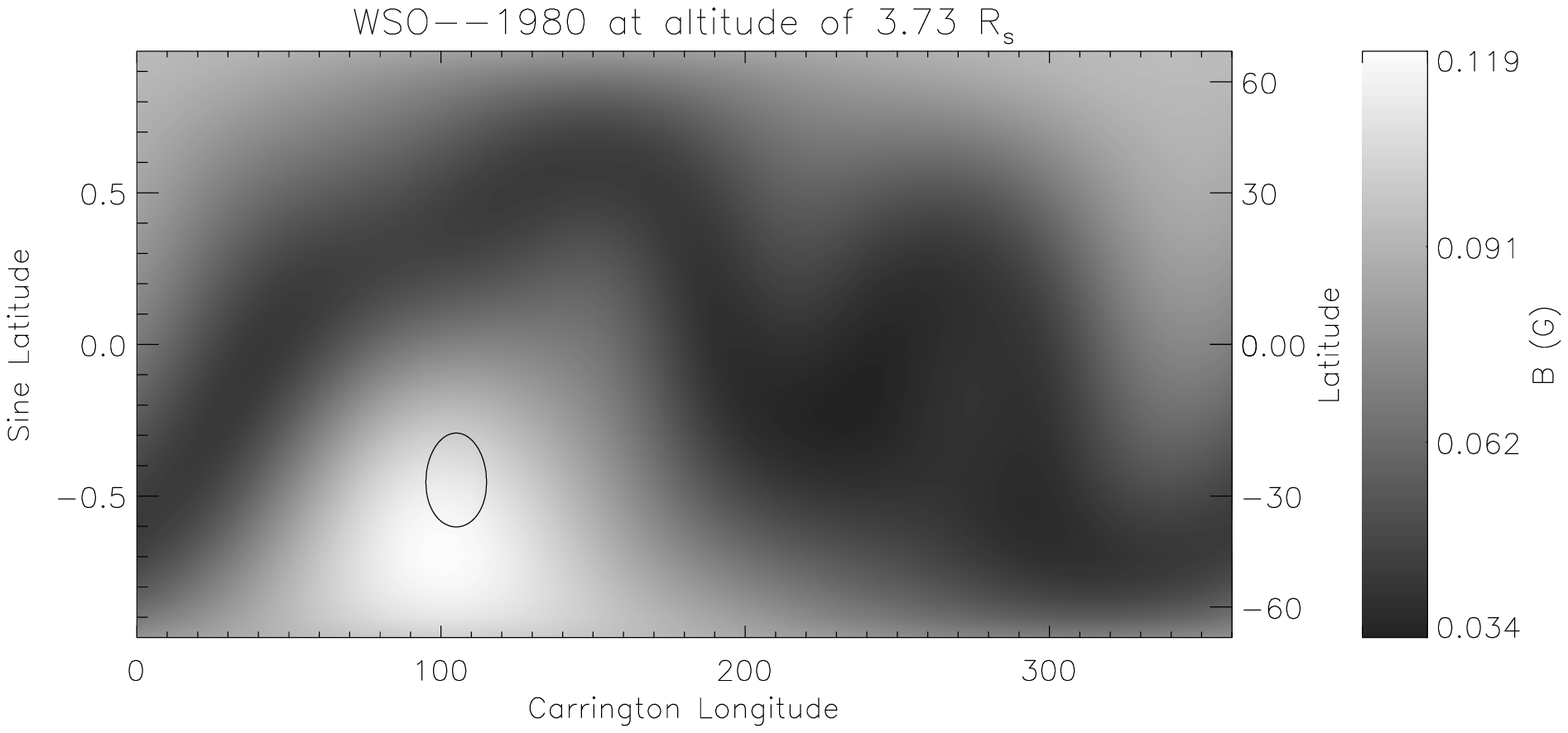}
\caption{The distribution of the extrapolated magnetic field strength at the height of 3.73 $R_s$, where the DH type II began to emit.}
\label{fg_1_4}
\end{center}
\end{figure*}

Following the instruction in Section 2, the coronal magnetic field extrapolation is
performed by using CSSS model based on the WSO synoptic chart of 
Carrington rotation 1980 corresponding to the period from 13:38 UT 2001 
August 23 to 19:51 UT 2001 September 19. Figure \ref{fg_1_4} shows the 
strength distribution of the extrapolated magnetic
field at the height of 3.73 $R_s$, where the DH type II just started.
It is found that the magnetic field strength varies in a large range from 0.03 to 0.12 G,
which indicates that the \va at the same altitude could be largely different
that directly influences the formation/strength of shocks.
The enclosed region right above the CME source region is treated as an effective
region, where the DH type II emitted. The mean strength of the magnetic
field at this height in this region is 0.11 G.

Combine the magnetic field strength with the plasma density,
the \va and fast-mode magnetosonic speed are calculated as
124$\sim$219 km/s and 180$\sim$219 km/s, respectively,
at the beginning of the DH type II. Therefore the shock strength,
indicated by the fast-mode magnetosonic Mach number,
of the CME is 3.43$\sim$4.18. The ranges of these values have
considered the 20\% error of the extrapolated magnetic field.
Usually a shock with Alfv\'{e}n Mach number larger than 2.7 is a
super-critical shock, like the bow shock ahead of the magnetosphere
of the Earth, while many interplanetary shocks are sub-critical ones.
The coronal shock driven by the CME on 2001 September 15
having a fast-mode magnetosonic Mach number larger than 3.43 means that,
although the shock driver was not so fast,
the shock was indeed strong and had capability to cause a strong DH type II,
which is factually observed by the Wind/Waves. 
Moreover, this strong shock also caused a major SEP event, in which
the maximum proton flux intensity with energy $>10$ MeV observed by GOES 
was 11.6 pfu. 

\subsection{2000 June 15 event}
\begin{figure*}[tb]
\begin{center}
\includegraphics[width=\hsize]{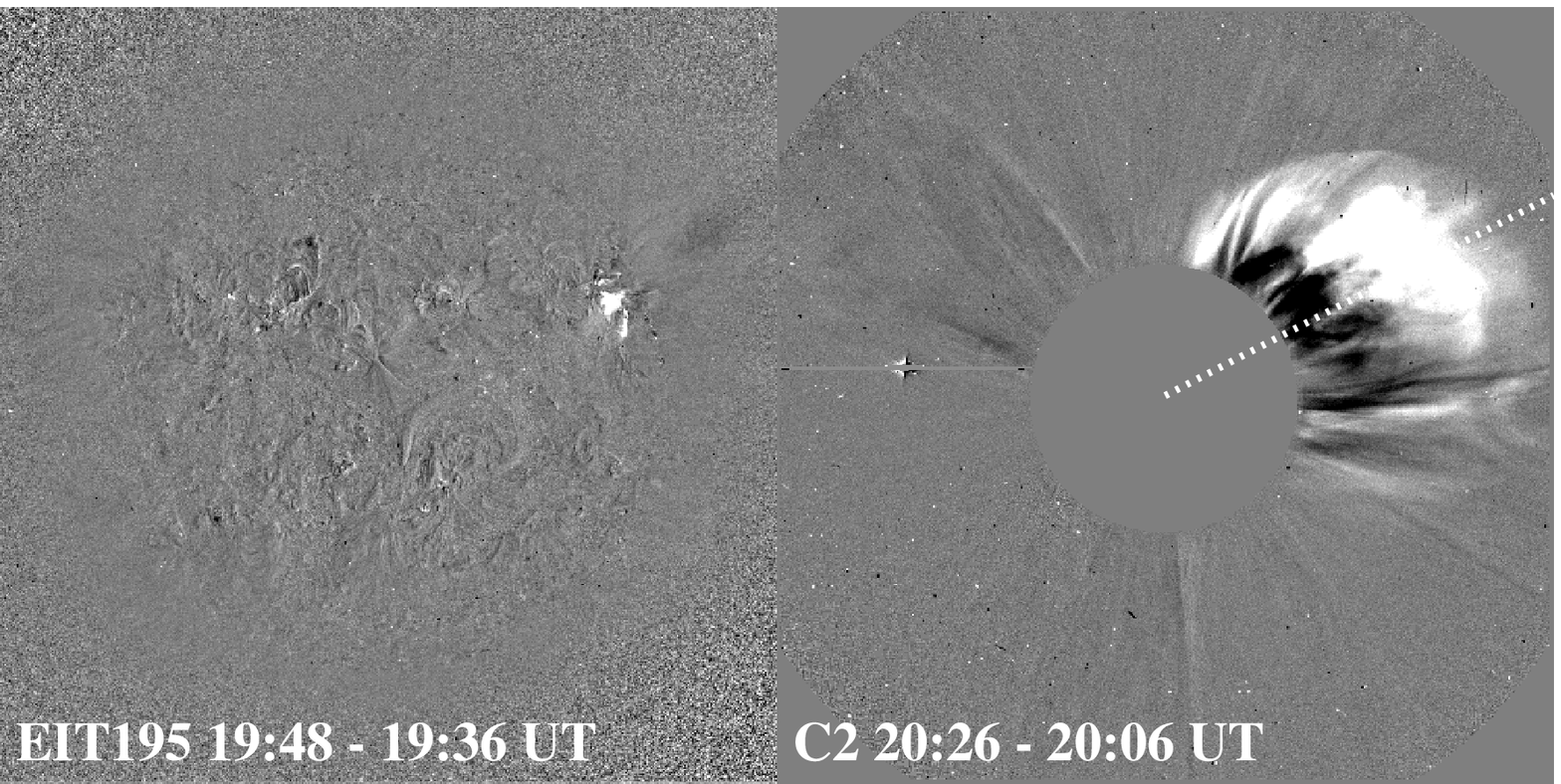}
\caption{The EIT and LASCO running-difference images of 2000 June 15 event.}
\label{fg_2_1}
\end{center}
\end{figure*}
\begin{figure*}[tb]
\begin{center}
\includegraphics[width=.9\hsize]{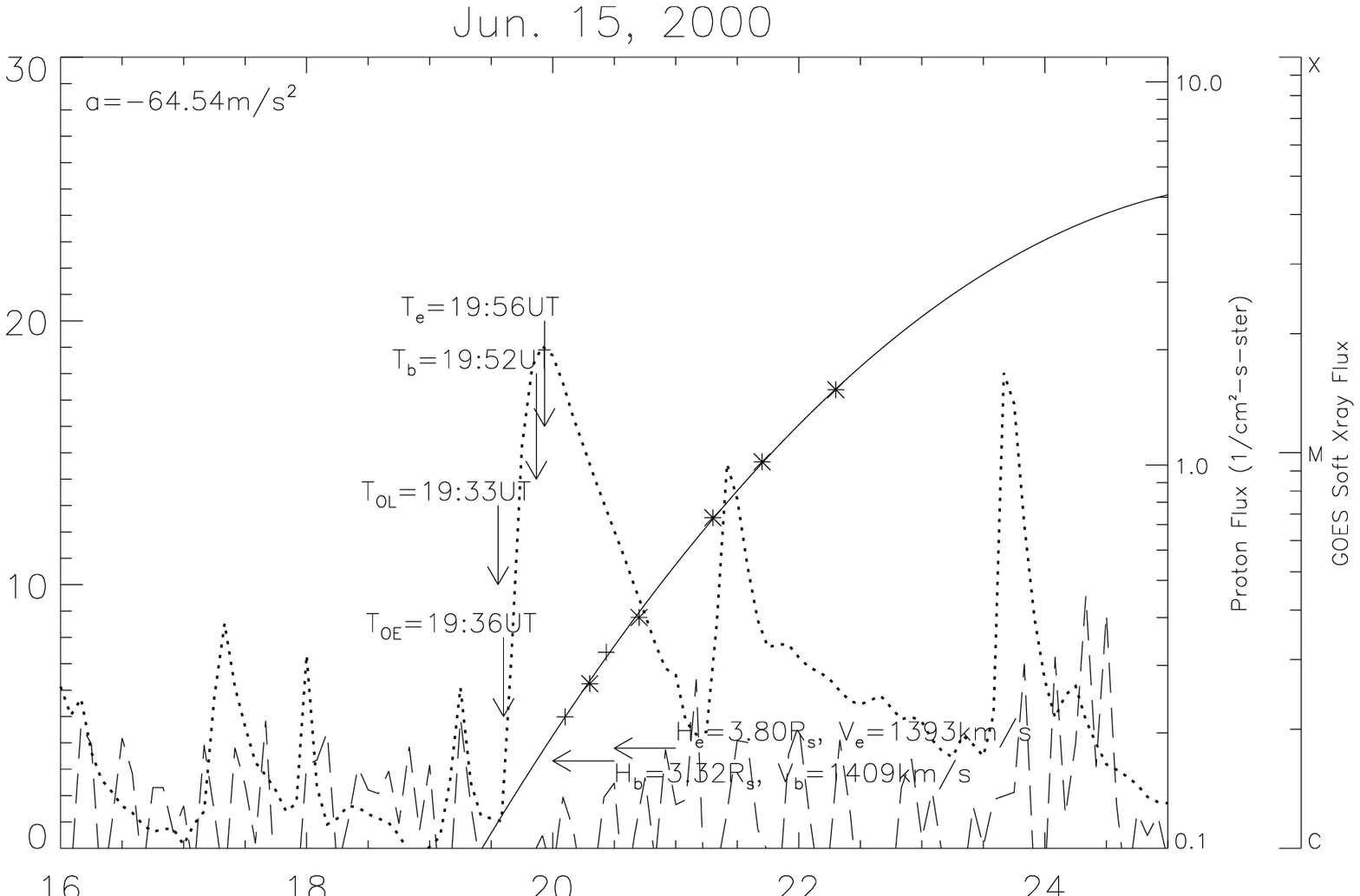}
\caption{The HT plot, proton flux profile and soft X-ray profile of the June 15, 2000 event.}
\label{fg_2_2}
\end{center}
\end{figure*}

The CME and its source region are presented in the
running-difference LASCO and EIT images of Figure \ref{fg_2_1}.
This is a fast CME first appearing at 20:06 UT on 2000 June 15
in LASCO/C2 and traveling along the northwest direction.
It originated from N19W56, where a sudden brightening could
be found. An M1.9 class flare associated with this event could be identified according to 
EIT images and the soft X-ray profile shown in Figure \ref{fg_2_2}.
The first eruptive signature recorded by EIT was at 19:36 UT.
Figure \ref{fg_2_2} shows the HT scatter of the leading front of this CME, 
in which all measured heights have been corrected to the heliocentric distance,
and the fitting curve by a second-order polynomial.
By extrapolating the curve down to the solar surface,
we find that the estimated onset time of the CME is 19:33 UT, 
close to the EIT onset time 19:36 UT.
This indicates that the fitted result is a good approximation to the reality.
The acceleration of this CME was $-64.54$ m/s$^{2}$,
and the speed was 1451 km/s at 2 $R_s$ and 1176 km/s at 10 $R_s$.

\begin{figure*}[tb]
\begin{center}
\includegraphics[width=.9\hsize]{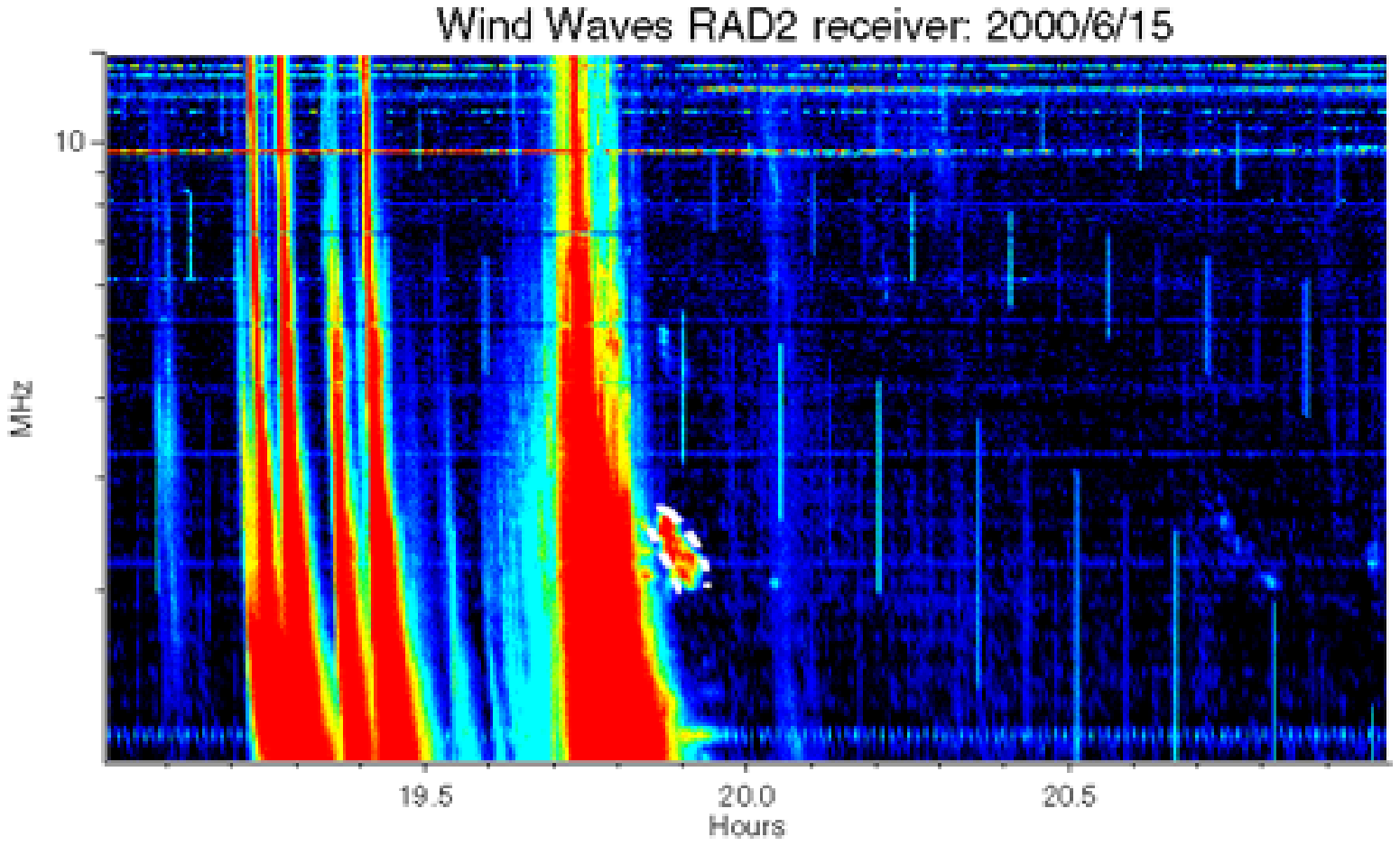}
\caption{The Wind/Waves observations during 19:00 -- 21:00 UT on June 15, 2000.}
\label{fg_2_3}
\end{center}
\end{figure*}

Figure \ref{fg_2_3} displays the Wind/Waves radio spectrum for this event.
A DH type II radio burst from about 19:52 to 19:56 UT is found. 
This radio burst is in the Wind/Waves Type II list
\footnote{http://lep694.gsfc.nasa.gov/waves/waves.html}.
This DH type II radio burst is shorter but has obvious fundamental and harmonic components. 
The fundamental began at about 19:52 UT around 2.4 MHz, 
and the harmonic appeared at about 5 MHz.
We could expect that the related CME-driven shock should be weaker than 
the shock driven by the first CME since the DH type II was weaker and shorter. 
The two dashed lines denote the upper and lower limits of the 
frequency of the fundamental component of this burst.
At the beginning $T_b=$19:52 UT, 
the frequency of fundamental component extended in the range from
 2.24 to 2.52 MHz approximately that corresponds to the plasma density
of 1.19$\sim$1.51 $\times 10^{-19}$ g/cm$^{3}$ according to Equation \ref{eq_rho}.
From the fitting curve in Figure \ref{fg_2_2}, the heliocentric distance
and speed of the CME are 3.32 $R_s$ and 1409 km/s at the beginning time,
$T_b$, of the DH type II, and 3.80 $R_s$ and 1393 km/s at the end time,
$T_e$. It was much faster than the 2001 September 15 CME.
\begin{figure*}[tb]
\begin{center}
\includegraphics[width=.9\hsize]{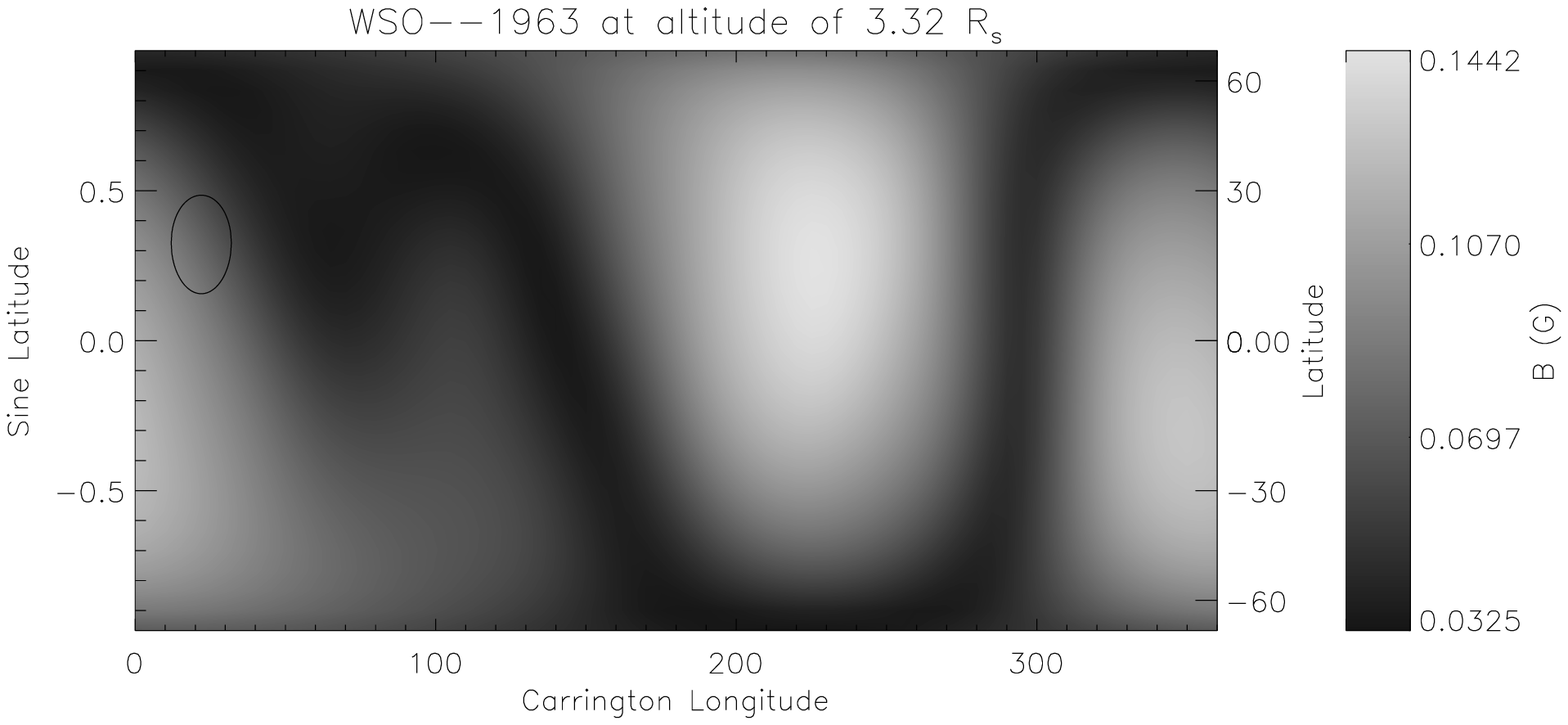}
\caption{The extrapolation magnetic field strength at the height of 3.32 $R_s$,
where the DH type II began to emit.}
\label{fg_2_4}
\end{center}
\end{figure*}

The WSO synoptic chart of Carrington rotation 1963,
which covers the period from 03:06 UT 2000 May 17 to 08:17 UT 2000 Jun 13,
is used to extrapolate the coronal magnetic filed.
We use the previous synoptic chart rather than the one right covering the CME onset time,
because the time of the CME source region passing across the central meridian is in  
Carrington rotation 1963. 
Figure \ref{fg_2_4} exhibits strength distribution of the
magnetic field at $H=3.32 R_s$, where the DH type II began to
emit. 
The magnetic filed strength varies from 0.033 to 0.144 G,
a larger range than that in the first event.
The effective region above the CME source region is enclosed by
the black circle in Figure \ref{fg_2_4}.
The mean magnetic field strength at this height in the effective region is 0.076G.

The calculation gives the values of \va and fast-mode magnetosonic speed for this event
439$\sim$741km/s and 439$\sim $741km/s, respectively, at
the beginning of the DH type II. Therefore the shock strength, $M_{ms}$,
of the CME is 1.90$\sim$3.21. The smaller Mach number 
indicates that the shock was relatively weak, as expected before.
The lower limit of $M_{ms}$, 1.90, indicates that the shock may be as weak as 
that just exceeds the critical condition (Mach number is 1.6) 
of a shock generating suprathermal electrons \citep{Mann_etal_1995}. 

\section{Comparison of the two cases with respect to the SEP flux}

Fast CMEs are usually thought to be a good producer of SEP events since they
usually drive a shock ahead \citep[e.g.,][]{Kahler_2001, Lario_2005}.
It is realized that the SEP events are not always fully observed near the Earth
even if a CME does generate them. This is because these energetic particles
generally move along the spiral interplanetary magnetic field lines,
and only the particles on the field lines connecting to the
Earth could be recorded. Thus those fast CMEs originating from western
hemisphere are thought to be more likelihood to cause an SEP event near the Earth.
As shown by \citet{Gopalswamy_etal_2004}, 
51 out of 60 SEP events are caused by CMEs originating from the western hemisphere. 

However, counter examples, that fast CMEs do not cause an SEP event
while slow CMEs do, do exist. We ever studied
56 fast (projected speed $\ge$ 1000 km/s) CMEs
originating from the western hemisphere during 1997--2003,
and found 12 CMEs did not cause any enhancement of
energetic proton flux at the proton energy $\ge$ 10 MeV \citep{Shen_etal_2006}. 
\citet{Kahler_Vourlidas_2005} presented 16 SEP-poor fast CMEs with speed large than 1060 km/s.  
\citet{Reinard_Andrews_2006} also found 14 CMEs with speed $\ge$ 1000 km/s 
did not cause an SEP event. A possible explanation is that the CME speed can not always
reflect the real shock strength. The comparison of flux profiles of protons with energy 
larger than 10 MeV between the two events analyzed here just illustrates this issue.
Of course, there is an alternative explanation for those slow events associated 
with SEPs, that the energetic particles do not come from shocks, but from something else like
flares. Such phenomenon will not be addressed in this paper.

\begin{table*}
    \centering \tiny \caption{Comparison of the 2001 September 15 event and 2000 June 15 event} \label{tb_1} \tabcolsep 3pt
    \begin{tabular}{cc|cc|ccc|c|ccccc|c}
No. &Date &\multicolumn{2}{c|}{DH Type IIs} &\multicolumn{3}{c|}{CMEs} &Flares&\multicolumn{5}{c|}{Derived Quantities} &SEP \\
& &$T_b$ &F &Loc &H &V & &$\rho$ &$B$ &$V_a$ &$V_{ms}$ &$M_{ms}$ & \\
& &UT &MHz & &$R_s$ &km/s &&$10^{-18}$ g/cm$^{3}$ &G &km/s &km/s & &pfu \\
    \hline
1 &2001/09/15 &11:51 &11.06$\sim$13.07 &S27W48 &3.73 &752 &M1.5&2.91$\sim$4.06&0.088$\sim$0.132 &124$\sim$219&180$\sim$219&3.43$\sim$4.18 &11.6\\
2 &2000/06/15 &21:02 &1.52$\sim$1.80 &N19W56 &3.32 &1409 &M1.9&0.119$\sim$0.151&0.061$\sim$0.091&439$\sim$741&439$\sim$741&1.90$\sim$3.21 & -- \\
    \end{tabular}
    \end{table*}

The two events (numbered 1 and 2, respectively) have
been summarized in Table \ref{tb_1} including the DH type II,
CME location and speed, flare intensity, plasma density, magnetic field strength,
\va, fast magnetosonic speed, and fast magnetosonic Mach number
as well as the SEP intensity. Both the CMEs originated
from the western hemisphere in the almost same longitude, 
and were associated with the similar flares which had the similar 
intensity (class of M1.5 and M1.9, respectively), 
so they are comparable.
The 2001 September 15 CME (number 1) is not a very fast CME.
Its speed is 752 km/s, about half of that of the 2000
June 15 CME (number 2). Based only on the CME speeds, we may
expect that CME 2 should produce a larger SEP event.
However, the fact is that CME 2 did not produce a major SEP event,
but CME 1 caused a major SEP event with the peak flux reaching 11.6 pfu
(as listed in the last column of Table \ref{tb_1}).

The flux profiles of the protons with energy larger than 10 MeV
recorded by GOES satellites have been presented as the dashed lines
in Figure \ref{fg_1_2} and \ref{fg_2_2}. The GOES instrument 
has a high background, but it will not affect our results, because 
GOES proton flux profiles are mainly used to determine whether or not there 
was an SEP event. Assuming zero scattering of protons propagating in 
the interplanetary space, the proton flight-time of about one hour 
from the Sun to 1 AU has been deducted from the proton flux profile.
For CME 1, the proton flux increased soon after the CME onset 
and the maximum value of this proton flux is 11.6 pfu. 
For CME 2, the proton flux profile fluctuates (comparable to the 
background noise) at the bottom of Figure
\ref{fg_2_2} by choosing the scale as the same as that in
Figure \ref{fg_1_2}, that suggests that there was no significant 
enhancement of the energetic proton flux. The observational facts are 
consistent with our calculated shock strengths,
which indicate that CME 1 drove a stronger shock though
it was relatively slow while CME 2 drove a relatively weak shock 
though it was very fast.

\section{Conclusion and Discussion}

In this study, we present an improved method to calculate the \va and
shock strength near the Sun ($> 2 R_s$),
in which as many observations as possible are used.
We use DH type II radio bursts instead of 1-D global density model to deduce
the ambient plasma density, use LASCO and EIT images to deduce
the heliocentric distance and shock speed at the time of DH type IIs emitting,
and use WSO synoptic charts and CSSS model to extrapolate the coronal magnetic field.
The plasma density, height and speed of shocks and magnetic field strength 
are all deduced independently.
Two events, a relatively slow CME on 2001 September 15 and a
very fast CME on 2000 June 15, are chosen to describe the
calculation process. The calculation results suggest
that the first event had a stronger shock with Mach number 
larger than 3.43 and up to 4.18, while the second event had a 
relatively weak shock with Mach number less than 3.21 and as low as 1.90. 
Here we would like to emphasize the lower limit of the Mach number
of the second shock, which is just above the critical point of
a shock generating superthermal electrons as suggested by \citet{Mann_etal_1995}.
These results are well consistent with the radio observations that a stronger and longer 
DH type II radio burst is found during the first event and a shorter DH 
type II during the second event.

The two events are further compared in respect of the capability
of SEP generation. It is found that the first CME produced a major SEP event
though it is relatively slow, while the second CME did not
cause a significant enhancement of energetic protons though it is very
fast. The fact also agrees with our calculations that the first
CME drove a strong shock while the second CME drove a weak shock.
The consistency in the comparisons of our calculations with both radio emissions and SEP
flux implies that the calculation method we proposed here is
reliable. According to the LASCO CME catalog\footnote{http://cdaw.gsfc.nasa.gov/CME\_list/}, 
the energy of the first CME was $\sim6.0\times 10^{30}$ erg and that of the 
second CME was $\sim3.1\times10^{31}$ erg. This means that neither 
speed nor kinematic energy could correctly reflect the real strength of 
potential shocks. The importance of the accurately estimation of shock
strength to the prediction of SEP events is therefore emphasized.

To see how large different between our calculations and those by
ideal models, Table \ref{tb_2} is compiled. For each event, the
first row shows our calculated parameters that have been shown in Table
\ref{tb_1}, and the second row lists the plasma density by one 
fold Newkirk model \citep{Newkirk_1961}, the magnetic field
strength by $B(r)=B_s(\frac{R_s}{r})^2$ with $B_s=2.2$ G, and the
corresponding fast-mode magnetosonic speed and Mach number. The
results by ideal models suggest that the first CME drove a
relatively weak shock and the second CME drove a stronger shock. 
This is opposite to our conclusion, and obviously contrary to 
the radio and SEP observations shown before.

\begin{table*}
   \centering \footnotesize \caption{Comparison of our results with ideal models} \label{tb_2} 
   \begin{tabular}{cccccccc}
No. &Date &H &V &$\rho$ &$B$ &$V_{ms}$ &$M_{ms}$  \\
&  &$R_s$ &km/s &$10^{-18}$ g/cm$^{3}$ &G &km/s &km/s  \\
\hline
1 &2001/09/15 &3.73 &752  & 2.91$\sim$4.06&0.088$\sim$0.132  &180$\sim$219 &3.43$\sim$4.18  \\
  &\multicolumn{3}{c}{Ideal Models}     &1.17        &0.16               &413           &1.82  \\
2 &2000/06/15 &3.32 &1409 &0.119$\sim$0.151 &0.061$\sim$0.091&439$\sim$741&1.90$\sim$3.21  \\
  &\multicolumn{3}{c}{Ideal Models}     &1.62        &0.20              &442           &3.19 \\
    \end{tabular}
    \end{table*}

Why the slower CME drove a stronger shock can be answered
according to Table \ref{tb_1}. The ambient plasma
ahead of CME 1 got based on the DH type II is much denser than that ahead of CME 2 while 
the heights of them are almost the same.
In ideal density models, the plasma density should decrease with increasing
height. However, although the height in event 1 is 3.73 $R_s$
higher than 3.32 $R_s$ in event 2, the
plasma density in event 1 is about 27 times of that in event 2. 
This is the main reason why the fast magnetosonic speed in event 
1 is much smaller than that in event 2. In addition, one may have noticed that 
the variation of density and magnetic field strength
with the time and/or location changing (even in the same altitude) is significant. 
The two studied events show that the density could vary over about one order, and 
the magnetic field strength could also vary over a few orders 
(Fig. \ref{fg_1_4} and \ref{fg_2_4}). Such large variations lead to 
the unreliability of using CME speed as a proxy of the shock strength.

SEP is an important effect of space weather. In this paper, we only regard the peak
intensity of energetic proton flux with energy larger than 10 MeV, one of the most 
concerned issue in space weather research/application. Actually, SEP acceleration and 
its manifestation are very complicated even in the events we studied here. Although 
there was no significant enhancement of the proton flux in event 2, the enhancement 
of the electron flux (up to $\sim10^5$ particles/cm$^{2}$-s-ster-MeV 
for energy range of 38--53 keV) observed by ACE/EPAM instrument
was much larger than that ($\sim2\times10^4$ particles/cm$^{2}$-s-ster-MeV) in event 1, 
and the enhancements of heavy ions in event 2 were also larger than those in event 1.
To fully address these phenomena needs the knowledge of the sources of these 
energetic particles (they may come from different places, shocks and/or flares), 
the priorities of species to be accelerated under different acceleration mechanisms, and
so on, which is beyond the scope of this paper.

Even so, how to predict the intensities of SEP events is still an interesting enough topic. 
Although it is well known that shock acceleration is the main mechanism of SEP generation, 
the correlation between shock strength and SEP intensity is still not well established. 
There is a general relationship between CME speed and SEP intensity; however
the scattering is much large \citep{Reames_2000}. 
A major reason is that the CME speed can not reflect the real strength 
of the CME-driven shock. Besides, based on our calculation, the presence of shocks seems 
not to be the sufficient condition of SEP generation, there may be a threshold 
in strength, only above which the shock is able to accelerate energetic 
protons efficiently. Thus, extending our present work to a statistical study 
that we are doing now should be of benefit to our understanding of the proton
acceleration at shock and space weather research.

\acknowledgments{
We acknowledge the use of the data from the LASCO and EIT instruments 
on board SOHO, Wind/Waves instrument, GOES satellites, ACE/EPAM instrument, and WSO.
SOHO is a project of international cooperation between ESA and NASA. 
We thank Drs. Yihua Yan and Shujuan Wang for the help 
in analyzing radio bursts. We are grateful to the referee's comments and suggestions.
This work is supported by the grants from the NSF of China
(40574063, 40525014, 40336052, 40404014), the 973 project (2006BC806304),
the Chinese Academy of Sciences (KZCX3-SW-144 and the startup fund),
and the Program for New Century Excellent Talents in University (NCET-04-0578).}


\begin{thebibliography}{55}
\expandafter\ifx\csname natexlab\endcsname\relax\def\natexlab#1{#1}\fi

\bibitem[{Altschuler \& Newkirk(1969)}]{Altschuler_Newkirk_1969}
Altschuler, M.~D., \& Newkirk. 1969, Sol. Phys., 9, 131

\bibitem[{Bougeret {et~al.}(1995)Bougeret, Kaiser, Kellogg, Manning, Goetz,
  Monson, Monge, Friel, Meetre, Perche, \& Sitruk}]{Bougeret_etal_1995}
Bougeret, J.-L., Kaiser, M.~L., Kellogg, P.~J., Manning, R., Goetz, K., Monson,
  S.~J., Monge, N., Friel, L., Meetre, C.~A., Perche, C., \& Sitruk,
  L.~andHoang, S. 1995, Space Sci. Rev., 71, 231

\bibitem[{Brueckner {et~al.}(1995)Brueckner, Howard, Koomen, Korendyke,
  Michels, Moses, Socker, Dere, Lamy, Llebaria, Bout, Schwenn, Simnett,
  Bedford, \& Eyles}]{Brueckner_etal_1995}
Brueckner, G.~E., Howard, R.~A., Koomen, M.~J., Korendyke, C.~M., Michels,
  D.~J., Moses, J.~D., Socker, D.~G., Dere, K.~P., Lamy, P.~L., Llebaria, A.,
  Bout, M.~V., Schwenn, R., Simnett, G.~M., Bedford, D.~K., \& Eyles, C.~J.
  1995, Sol. Phys., 162, 357

\bibitem[{Cane \& Lario(2006)}]{Cane_Lario_2006}
Cane, H.~V., \& Lario, D. 2006, Space Sci. Rev., 123, 45

\bibitem[{Cane {et~al.}(1981)Cane, Stone, Fainberg, Steinberg, Hoang, \&
  Stewart}]{Cane_etal_1981}
Cane, H.~V., Stone, R.~G., Fainberg, J., Steinberg, J.~L., Hoang, S., \&
  Stewart, R.~T. 1981, Geoph. Phy. Lett., 8, 1285

\bibitem[{Cho {et~al.}(2007)Cho, Lee, Moon, Dryer, Bong, Kim, \&
  Park}]{Cho_etal_2007}
Cho, K.-S., Lee, J., Moon, Y.-J., Dryer, M., Bong, S.-C., Kim, Y.-H., \& Park,
  Y.~D. 2007, Astron. \& Astrophys., 461, 1121

\bibitem[{Cho {et~al.}(2005)Cho, Moon, Dryer, Shanmugaraju, Fry, Kim, Bong, \&
  Park}]{Cho_etal_2005}
Cho, K.~S., Moon, Y.~J., Dryer, M., Shanmugaraju, A., Fry, C.~D., Kim, Y.~H.,
  Bong, S.~C., \& Park, Y.~D. 2005, J. Geophys. Res., 110, A12101

\bibitem[{Ciaravella {et~al.}(2006)Ciaravella, Raymond, \&
  Kahler}]{Ciaravella_etal_2006}
Ciaravella, A., Raymond, J.~C., \& Kahler, S.~W. 2006, Astrophys. J., 652, 774

\bibitem[{Ciaravella {et~al.}(2005)Ciaravella, Raymond, Kahler, Vourlidas, \&
  Li}]{Ciaravella_etal_2005}
Ciaravella, A., Raymond, J.~C., Kahler, S.~W., Vourlidas, A., \& Li, J. 2005,
  Astrophys. J., 621, 1121

\bibitem[{Cliver {et~al.}(2004)Cliver, Kahler, \& Reames}]{Cliver_etal_2004}
Cliver, E.~W., Kahler, S.~W., \& Reames, D.~V. 2004, Astrophy. J, 605, 902

\bibitem[{Delaboudiniere {et~al.}(1995)Delaboudiniere, Artzner, Brunaud,
  Gabriel, Hochedez, Millier, Song, Au, Dere, Howard, Kreplin, Michels, Moses,
  Defise, Jamar, Rochus, Chauvineau, Marioge, Catura, Lemen, Shing, Stern,
  Gurman, Neupert, Maucherat, Clette, Cugnon, \& van
  Dessel}]{Delaboudiniere_etal_1995}
Delaboudiniere, J.-P., Artzner, G.~E., Brunaud, J., Gabriel, A.~H., Hochedez,
  J.~F., Millier, F., Song, X.~Y., Au, B., Dere, K.~P., Howard, R.~A., Kreplin,
  R., Michels, D.~J., Moses, J.~D., Defise, J.~M., Jamar, C., Rochus, P.,
  Chauvineau, J.~P., Marioge, J.~P., Catura, R.~C., Lemen, J.~R., Shing, L.,
  Stern, R.~A., Gurman, J.~B., Neupert, W.~M., Maucherat, A., Clette, F.,
  Cugnon, P., \& van Dessel, E.~L. 1995, Sol. Phys., 162, 291

\bibitem[{Dulk {et~al.}(1971)Dulk, Altschuler, \& Smerd}]{Dulk_etal_1971}
Dulk, G.~A., Altschuler, M.~D., \& Smerd, S.~F. 1971, Astrophys. J., 8, L235

\bibitem[{Gary {et~al.}(1984)Gary, Dulk, House, Illing, Sawyer, Wagner, McLean,
  \& Hildner}]{Gary_etal_1984}
Gary, D.~E., Dulk, G.~A., House, L., Illing, R., Sawyer, C., Wagner, W.~J.,
  McLean, D.~J., \& Hildner, E. 1984, Astron. \& Astrophys., 134, 222

\bibitem[{Gopalswamy {et~al.}(2005)Gopalswamy, E., Yashiro, Nunes, Kaiser, \&
  Howard}]{Gopalswamy_etal_2005}
Gopalswamy, N., E., A.-R., Yashiro, S., Nunes, S., Kaiser, M.~L., \& Howard,
  R.~A. 2005, J. Geophys. Res., 110, A12S07

\bibitem[{Gopalswamy \& Kaiser(2002)}]{Gopalswamy_Kaiser_2002}
Gopalswamy, N., \& Kaiser, M.~L. 2002, Adv. Space Res., 29, 307

\bibitem[{Gopalswamy {et~al.}(1997)Gopalswamy, Kundu, Manoharan, Raoult, Nitta,
  \& Zarka}]{Gopalswamy_etal_1997}
Gopalswamy, N., Kundu, M.~R., Manoharan, P.~K., Raoult, A., Nitta, N., \&
  Zarka, P. 1997, Astrophys. J., 1036

\bibitem[{Gopalswamy {et~al.}(2001)Gopalswamy, Yashiro, Krucker, Howard, \&
  L}]{Gopalswamy_etal_2001a}
Gopalswamy, N., Yashiro, S., Krucker, S., Howard, R.~A., \& L, B.~J. 2001, J.
  Geophys. Res., 106, 29219

\bibitem[{Gopalswamy {et~al.}(2004)Gopalswamy, Yashiro, Krucker, Stenborg, \&
  Howard}]{Gopalswamy_etal_2004}
Gopalswamy, N., Yashiro, S., Krucker, S., Stenborg, G., \& Howard, R.~A. 2004,
  J. Geophys. Res., 109, A12105

\bibitem[{Holman \& Pesses(1983)}]{Holman_Pesses_1983}
Holman, G.~D., \& Pesses, M.~E. 1983, Astrophys. J., 267, 837

\bibitem[{Jones \& Ellison(1991)}]{Jones_Ellison_1991}
Jones, F.~C., \& Ellison, D.~C. 1991, Space Sci. Rev., 58, 259

\bibitem[{Kahler(1994)}]{Kahler_1994}
Kahler, S.~W. 1994, Astrophys. J., 428, 837

\bibitem[{Kahler(2001)}]{Kahler_2001}
---. 2001, J. Geophys. Res., 106, 20947

\bibitem[{Kahler \& Vourlidas(2005)}]{Kahler_Vourlidas_2005}
Kahler, S.~W., \& Vourlidas, A. 2005, J. Geophys. Res., 110, A12S01

\bibitem[{Klein {et~al.}(1999)Klein, Chupp, Trottet, Magun, Dunphy, Rieger, \&
  Urpo}]{Klein_etal_1999}
Klein, K.-L., Chupp, E.~L., Trottet, G., Magun, A., Dunphy, P.~P., Rieger, E.,
  \& Urpo, S. 1999, Astron. \& Astrophys., 348, 271

\bibitem[{Kohl {et~al.}(1995)Kohl, Esser, Gardner, Habbal, Daigneau, Dennis,
  Nystrom, Panasyuk, Raymond, Smith, Strachan, van Ballegooijen, Noci,
  Fineschi, Romoli, Ciaravella, Modigliani, Huber, Antonucci, Benna, Giordano,
  Tondello, Nicolosi, Naletto, Pernechele, Spadaro, Poletto, Livi, von~der
  Luhe, Geiss, Timothy, Gloeckler, Basile, Brusa, Wood, Siegmund, Fowler,
  Fisher, \& Jhabvala}]{Kohl_etal_1995}
Kohl, J.~L., Esser, R., Gardner, L.~D., Habbal, S., Daigneau, P.~S., Dennis,
  E.~F., Nystrom, G.~U., Panasyuk, A., Raymond, J.~C., Smith, P.~L., Strachan,
  L., van Ballegooijen, A.~A., Noci, G., Fineschi, S., Romoli, M., Ciaravella,
  A., Modigliani, A., Huber, M. C.~E., Antonucci, E., Benna, C., Giordano, S.,
  Tondello, G., Nicolosi, P., Naletto, G., Pernechele, C., Spadaro, D.,
  Poletto, G., Livi, S., von~der Luhe, O., Geiss, J., Timothy, J.~G.,
  Gloeckler, G. adn~Allegra, A., Basile, G., Brusa, R., Wood, B., Siegmund, O.
  H.~W., Fowler, W., Fisher, R., \& Jhabvala, M. 1995, Sol. Phys., 162, 313

\bibitem[{Lario(2005)}]{Lario_2005}
Lario, D. 2005, Adv. Space Res., 36, 2279

\bibitem[{Maia {et~al.}(2000)Maia, Pick, Vourlidas, \& Howard}]{Maia_etal_2000}
Maia, D., Pick, M., Vourlidas, A., \& Howard, R. 2000, Astrophys. J., 528, L49

\bibitem[{Malitson {et~al.}(1973)Malitson, Fainberg, \&
  Stone}]{Malitson_etal_1973}
Malitson, H.~H., Fainberg, J., \& Stone, R.~G. 1973, Astrophys. Lett., 14, 111

\bibitem[{Mancuso {et~al.}(2002)Mancuso, Raymond, Kohl, Ko, Uzzo, \&
  Wu}]{Mancuso_etal_2002}
Mancuso, S., Raymond, J.~C., Kohl, J., Ko, Y.~K., Uzzo, M., \& Wu, R. 2002,
  Astron. \& Astrophys, 383, 267

\bibitem[{Mann {et~al.}(1995)Mann, Classen, \& Aurass}]{Mann_etal_1995}
Mann, G., Classen, T., \& Aurass, H. 1995, Astron. \& Astrophys, 295, 775

\bibitem[{Mann {et~al.}(2003)Mann, Klassen, Aurass, \&
  Classen}]{Mann_etal_2003}
Mann, G., Klassen, A., Aurass, H., \& Classen, H.~T. 2003, Astron. \&
  Astrophys, 400, 329

\bibitem[{Nelson \& Melrose(1985)}]{Nelson_Melrose_1985}
Nelson, G.~J., \& Melrose, D.~B. 1985, in Solar radiophysics: Studies of
  emission from the sun at metre wavelengths, ed. D.~C. McLean \& N.~R. Labrum
  (Cambridge University Press), 333--359

\bibitem[{Newkirk(1961)}]{Newkirk_1961}
Newkirk, Gordon, J. 1961, Astrophys. J, 133, 983

\bibitem[{Priest(1982)}]{Priest_1982}
Priest, E.~R. 1982, Solar Magnetohydrodynamics (Dordrecht:Reidel)

\bibitem[{Raouafi {et~al.}(2004)Raouafi, Mancuso, Solanki, Inhester, Mierla,
  Stenborg, Delaboudini¨¨re, \& Benna}]{Raouafi_etal_2004}
Raouafi, N.-E., Mancuso, S., Solanki, S.~K., Inhester, B., Mierla, M.,
  Stenborg, G., Delaboudini¨¨re, J.~P., \& Benna, C. 2004, Astron. \&
  Astrophys., 424, 1039

\bibitem[{Raymond {et~al.}(2000)Raymond, Barbara, Cyr, N., Kahler, Kaiser,
  Lara, Romoli, \& O'neal}]{Raymond_etal_2000}
Raymond, J.~C., Barbara, J.~T., Cyr, O. C.~S., N., G., Kahler, S., Kaiser, M.,
  Lara, A., Romoli, M., \& O'neal, R. 2000, Geophys. Res. Lett., 27, 1439

\bibitem[{Reames(1999)}]{Reames_1999}
Reames, D.~V. 1999, Space Sci. Rev., 90, 413

\bibitem[{Reames(2000)}]{Reames_2000}
Reames, D.~V. 2000, in AIP Conf. Proc., Vol. 516, Invited, Rapporteur, and
  Highlight Papers, ed. B.~L. Dingus, D.~B. Kieda, \& M.~H. Salamon, 289--300

\bibitem[{Reinard \& Andrews(2006)}]{Reinard_Andrews_2006}
Reinard, A.~A., \& Andrews, M.~A. 2006, Adv. Space Res., 38, 480

\bibitem[{Robinson \& Stewart(1985)}]{Robinson_Stewart_1985}
Robinson, R.~D., \& Stewart, R.~T. 1985, Sol. Phys., 97, 145

\bibitem[{Saito {et~al.}(1977)Saito, Poland, \& Munro}]{Saito_etal_1977}
Saito, K., Poland, A.~I., \& Munro, R.~H. 1977, Sol. Phys., 55, 121

\bibitem[{Schatten(1971)}]{Schatten_1971}
Schatten, K.~H. 1971, Cosmic Eletrodyn., 2, 232

\bibitem[{Schatten {et~al.}(1969)Schatten, W., \& F.}]{Schatten_etal_1969}
Schatten, K.~H., W., W.~J., \& F., N.~N. 1969, Solar Phys., 6, 442

\bibitem[{Schrijver {et~al.}(2005)Schrijver, DeRosa, Title, \&
  Metcalf}]{Schrijver_etal_2005}
Schrijver, C.~J., DeRosa, M.~L., Title, A.~M., \& Metcalf, T.~R. 2005,
  Astrophys. J., 628, 501

\bibitem[{Sheeley {et~al.}(1997)Sheeley, Wang, Hawley, Brueckner, Dere, Howard,
  Koomen, Korendyke, Michels, Paswaters, Socker, St.~Cyr, Wang, Lamy, Llebaria,
  Schwenn, Simnett, Plunkett, \& Biesecker}]{Sheeley_etal_1997}
Sheeley, N.~R., J., Wang, Y.-M., Hawley, S.~H., Brueckner, G.~E., Dere, K.~P.,
  Howard, R.~A., Koomen, M.~J., Korendyke, C.~M., Michels, D.~J., Paswaters,
  S.~E., Socker, D.~G., St.~Cyr, O.~C., Wang, D., Lamy, P.~L., Llebaria, A.,
  Schwenn, R., Simnett, G.~M., Plunkett, S., \& Biesecker, D.~A. 1997,
  Astrophys. J., 484, 472

\bibitem[{Shen {et~al.}(2006)Shen, Wang, Ye, \& Wang}]{Shen_etal_2006}
Shen, C.~l., Wang, Y.~M., Ye, P.~Z., \& Wang, S. 2006, Astrophys. J., 639, 510

\bibitem[{Vourlidas {et~al.}(2003)Vourlidas, Wu, Wang, Subramanian, \&
  Howard}]{Vourlidas_etal_2003}
Vourlidas, A., Wu, S.~T., Wang, A.~H., Subramanian, P., \& Howard, R.~A. 2003,
  Astrophys. J., 598, 1392

\bibitem[{Vr\v{s}nak {et~al.}(2004)Vr\v{s}nak, Magdalenic, \&
  Zlobec}]{Vrsnak_etal_2004}
Vr\v{s}nak, B., Magdalenic, J., \& Zlobec, P. 2004, Astron. \& Astrophys., 413,
  753

\bibitem[{Warmuth \& Mann(2005)}]{Warmuth_Mann_2005}
Warmuth, A., \& Mann, G. 2005, A\&A, 435, 1123

\bibitem[{Wild \& McCready(1950)}]{Wild_McCready_1950}
Wild, J.~P., \& McCready, L.~L. 1950, Austral. J. Sci. Res., A3, 387

\bibitem[{Zhao \& Hoeksema(1992)}]{Zhao_Hoeksema_1992}
Zhao, X.~P., \& Hoeksema, J.~T. 1992, Eur. Space Agency Spec. Publ., ESA
  SP-348, 117

\bibitem[{Zhao \& Hoeksema(1994{\natexlab{a}})}]{Zhao_Hoeksema_1994b}
---. 1994{\natexlab{a}}, Solar Phys., 151, 91

\bibitem[{Zhao \& Hoeksema(1994{\natexlab{b}})}]{Zhao_Hoeksema_1994a}
---. 1994{\natexlab{b}}, Adv. Space Res., COSPAR 94, 11

\bibitem[{Zhao \& Hoeksema(1995)}]{Zhao_Hoeksema_1995}
---. 1995, J. Geophys. Res., 100, 19

\bibitem[{Zhao {et~al.}(2002)Zhao, Hoeksema, \& Rich}]{Zhao_etal_2002a}
Zhao, X.~P., Hoeksema, J.~T., \& Rich, N.~B. 2002, Adv. Space Res., 29, 411

\end{thebibliography}
\end{document}